\documentclass[12pt,a4paper,dvips]{article}
\usepackage{a4p}
\usepackage{cite,mcite}
\usepackage{graphicx}
\usepackage{physics }
\usepackage{l3_title,ifthen}
\usepackage{rotating}
\usepackage{epsfig}

\journalname{Phys. Lett. B}
\date{26 January 2001}

\preprint{2001-008}

\usepackage{Lep}
\Lep{2}

\newcommand{\Zggto}{\ensuremath{\rm Z\gamma\gamma \rightarrow}~}
\newcommand{\eeto}{\ensuremath{\rm e^+e^- \rightarrow}~}

\newcommand{\qqgg}{\ensuremath{\rm q\bar{q}\gamma\gamma}}
\newcommand{\Zgg}{\ensuremath{\rm Z\gamma\gamma }~}

\newlength{\capindent}
\newlength{\capwidth}
\newlength{\figwidth}
\newcommand{\icaption}[2][!*!,!]{\hspace*{\capindent}%
  \begin{minipage}{\capwidth}
    \ifthenelse{\equal{#1}{!*!,!}}%
      {\caption{#2}}%
      {\caption[#1]{#2}}
  \end{minipage}}
\begin{document}
\setlength{\unitlength}{1mm}
\begin{titlepage}
\title{Study  of  the \boldmath{${\epem\ra\Zo\gamma\gamma}\rightarrow \qqgg $}
Process at LEP}
\author{The L3 Collaboration}
\begin{abstract}

 The process
$\epem\ra\Zo\gamma\gamma\rightarrow\qqgg $  is studied in   0.5\,fb$^{-1}$ 
of data
collected with the L3 detector 
at centre-of-mass energies between 
 $130.1\GeV$ and $201.7\GeV$.
Cross sections are measured and found to be  consistent  with the Standard Model expectations.
The study of the least energetic photon  constrains the 
 quartic gauge boson couplings to
$-0.008\GeV^{-2}  < a_0/\Lambda^2 < 0.005\GeV^{-2}$ and
$-0.007\GeV^{-2}  < a_c/\Lambda^2 < 0.011\GeV^{-2}$,  at
95\% confidence level. 

\end{abstract}

\submitted

\end{titlepage}

%
%

\section{Introduction}

The LEP data offer  new
insight into the Standard Model  of electroweak
interactions~\cite{sm_glashow} by investigating  
the production of three gauge bosons. Results were recently 
reported on studies of  
the reactions  $\epem\ra\Zo\gamma\gamma$ \cite{l3pap} and
$\epem\ra\Wp\Wm\gamma$~\cite{l3wwg,opalwwg}. 
This letter describes the extension of the study of the \eeto \Zgg process
 to centre-of-mass energies, $\sqrt{s}$, between 130 and 202 \GeV. 
Final states with hadrons and isolated photons are considered to select $\Zggto \qqgg$ 
events.

\begin{figure}[p]
  \begin{center}
      \mbox{\includegraphics[width=\figwidth]{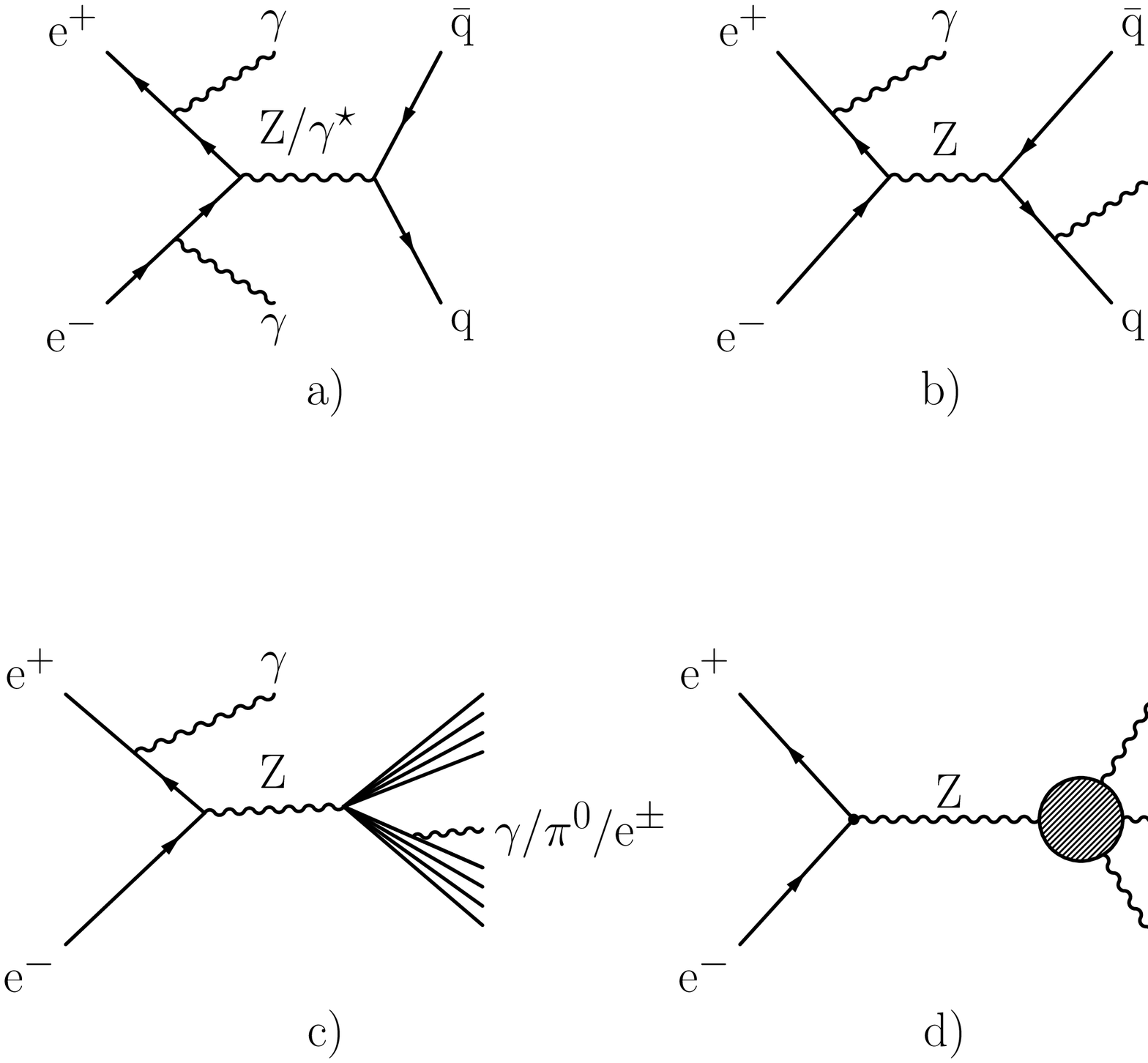}}\vspace{4cm}
      \icaption{Diagrams of a) the Standard Model 
contribution to \eeto \Zgg signal and  ``non-resonant'' background,
 b) the background  
from direct radiation of photon from the quarks,
 c) the background from photons,  misidentified electrons 
or unresolved $\pi^0$s originating from hadrons and
 d) the  anomalous QGC diagram. 
     \label{fig:0}}
  \end{center}
\end{figure}

In the Standard Model, the \eeto \Zgg  process occurs via  radiation of 
 photons from the incoming electron and/or positron. One 
possible diagram is presented in Figure~\ref{fig:0}a.

 The $\epem\ra\Zo\gamma\gamma$ signal is defined by  
phase-space requirements on the  energies $E_{\gamma}$ and angles $\theta_\gamma$ of the two photons,
 and on the propagator  mass $\sqrt{s'}$:
\begin{equation}
E_{\gamma}> 5 \GeV
\label{eq:sign1}
\end{equation}
\begin{equation}
|\cos{\theta_{\gamma}}|< 0.97
\label{eq:sign2}
\end{equation}
\begin{equation}
|\sqrt{s'}-m_{\rm Z}|< 2 \Gamma_{\rm Z}  
\label{eq:sign3}
\end{equation}
where $m_{\rm Z}$ and $\Gamma_{\rm Z}$ are the Z boson mass and
width.
In the following,  hadronic decays of the Z boson are considered. 
Events with hadrons and initial state photons falling outside the signal
definition cuts are referred to as ``non-resonant'' background. 

A single initial state radiation photon can also lower 
the effective centre-of-mass energy
of the $\epem$ collision to $m_{\rm Z}$, with the subsequent production of a
quark-antiquark pair. This photon can be mistaken for  the most
energetic photon of the $\epem\ra\Zo\gamma\gamma\ra\qqbar\gamma\gamma$
process.
Two sources  can then mimic the least energetic
photon: either 
the direct radiation of photons from the quarks or
photons originating from hadronic decays, misidentified electrons 
or unresolved $\pi^0$s.
These background processes  are   depicted  in Figures~\ref{fig:0}b and~\ref{fig:0}c, respectively.

In order to compare experimental results with $\eeto \qqgg$ matrix element
calculations, a further requirement is applied on the angle $\theta_{\rm
  \gamma q}$ between the photons and the nearest quark:
\begin{equation}
\cos{\theta_{\rm \gamma q}}< 0.98.
\label{eq:sign4}
\end{equation}
This cut  avoids collinear divergences. Its inclusion makes  the 
signal definition used here different from the previous one~\cite{l3pap}.
Signal 
cross sections calculated with the  KK2f Monte Carlo program\cite{KK2f}
range from 0.9\,pb at $\sqrt{s} = 130.1 \GeV$  
down to 0.3\,pb at $\sqrt{s} = 201.7 \GeV$.

The \Zgg final state could also originate from  the  $s$-channel exchange
of a Z boson, as presented in  Figure~\ref{fig:0}d.
This process is  forbidden at tree level in the Standard Model, but 
it is expected to occur in the presence of Quartic  Gauge boson Couplings 
(QGC) beyond the Standard Model.

\section{Data and Monte Carlo Samples}

This measurement uses  
data  collected with the L3
detector~\cite{l3_00} at
LEP in the years from  1995 through 1999, at centre-of-mass energies between
$\sqrt{s}=130.1\GeV$
and $\sqrt{s}=201.7\GeV$, for a total  integrated luminosity of
0.5\,fb$^{-1}$.  The centre-of-mass energies and
 the corresponding integrated luminosities
 are listed in Table \ref{tab:encm_lumi}.
 Given their relatively low luminosities, the
 $\sqrt{s} = 130.1 \GeV$ and $\sqrt{s} = 136.1 \GeV$
 data sample are combined into a single luminosity
 averaged sample at $\sqrt{s} = 133.1 \GeV$.
 Similarly  the $\sqrt{s} = 161.3 \GeV$
 and $\sqrt{s} = 172.3 \GeV$ samples are merged 
into a  single sample at $\sqrt{s} = 166.8 \GeV$.    

The KK2f Monte Carlo program 
is used to generate \eeto ${\rm {q\bar{q}(\gamma\gamma)}}$ events,
 that are assigned  to the 
 signal or the background according 
to the criteria (1)$-$(\ref{eq:sign4}).
The hadronisation process is   simulated
with the JETSET~\cite{pythia} program. 
Other  background processes are 
generated with the Monte Carlo programs PYTHIA \cite{pythia}
($\rm e^+ e^- \rightarrow Z \epem$ and $\rm e^+ e^- 
\rightarrow ZZ$), 
KORALZ\,~\cite{koralz} ($\rm e^+ e^- \rightarrow \tau^+ \tau^-
(\gamma)$), PHOJET\,~\cite{phojet} ($\rm e^+ e^- \rightarrow e^+ e^-$
hadrons) and  KORALW\,~\cite{koralw} for  $\Wp\Wm$ production  
except for the $\rm e\nu_\e q\bar q'$ final states,  generated
with EXCALIBUR~\cite{exca}.

The L3 detector response is simulated using the GEANT~\cite{geant} 
and GHEISHA~\cite{gheisha} 
programs, which model the effects of energy loss,
multiple scattering and showering in the detector. 
Time dependent detector inefficiencies, as monitored during data taking
periods, are also  simulated. 
\begin{table}[ht]
  \begin{center}
        \begin{tabular}{|c|c|r|}
       \hline
        $\sqrt{s}$ &  Integrated Luminosity \\
        $({\rm GeV})$ & (${\rm pb^{-1}}$) \\
       \hline
       133.1  &  \,\,\,12.0  \\
       166.8  & \,\,\,21.1 \\
       182.7  & \,\,\,55.3  \\ 
       188.7  & 176.3  \\
       191.6  & \,\,\,29.4 \\
       195.5  & \,\,\,83.7 \\
       199.5  & \,\,\,82.8 \\
       201.7  &  \,\,\,37.0 \\ 
        \hline
    \end{tabular}
    \end{center}
\caption{Average centre-of-mass energies and
 corresponding integrated luminosities of the data samples used for this analysis.}
\label{tab:encm_lumi}
\end{table}

\section{Event Selection}

The $\epem\ra\Zo\gamma\gamma\ra\qqbar\gamma\gamma$ selection demands  
 balanced
hadronic events with two isolated photons and small 
energy deposition at low polar angle.
Selection criteria on  photon energies and angles follow
 directly from the signal definition as $E_\gamma > 5\GeV$ and
 $|\cos{\theta_\gamma}|<0.97$.  
The invariant mass $M_{\rm q\bar{q}}$
of the reconstructed
 hadronic system, forced into two jets using 
 the DURHAM algorithm~\cite{durham}, is required to be consistent 
with a Z boson decaying into hadrons, $72\GeV < M_{\rm q\bar{q}}< 116\GeV$.

 The main background after these  requirements is 
 due to the ``non-resonant'' production of two 
photons and a hadronic system. The relativistic velocity
 $\beta_{\rm Z} = p_{\rm Z}/E_{\rm Z}$ of the system recoiling against  the
photons, calculated assuming 
its mass to be the nominal Z mass, is  larger for part of these background
events than for the signal and   an upper cut  
is used to reject those events.
 It  is  optimised for each centre-of-mass energy, as listed in
 Table~\ref{sel_cuts}.

\begin{figure}[p]
  \begin{center}
    \begin{tabular}{cc}
      \mbox{\includegraphics[width=.5\figwidth]{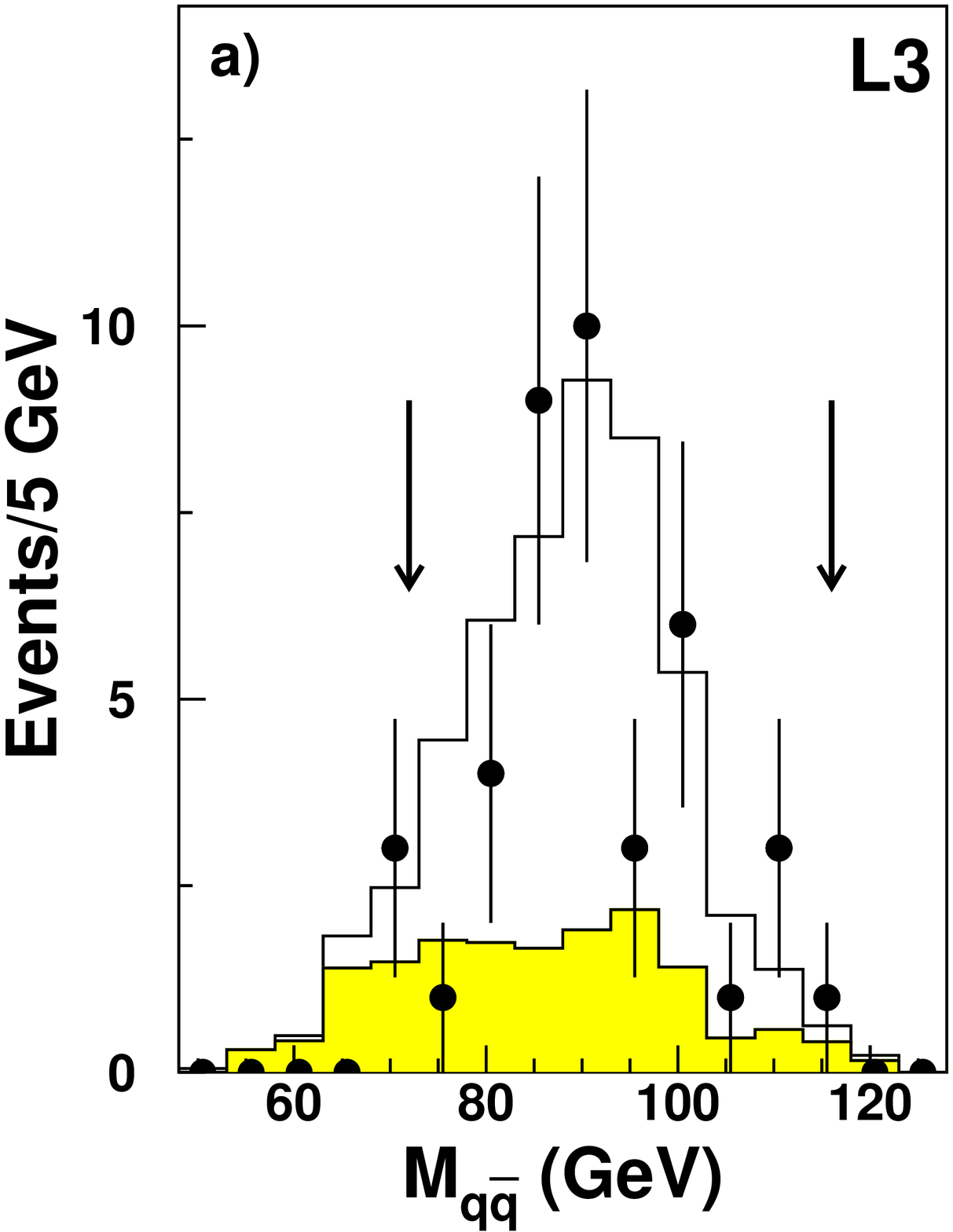}} &
      \mbox{\includegraphics[width=.5\figwidth]{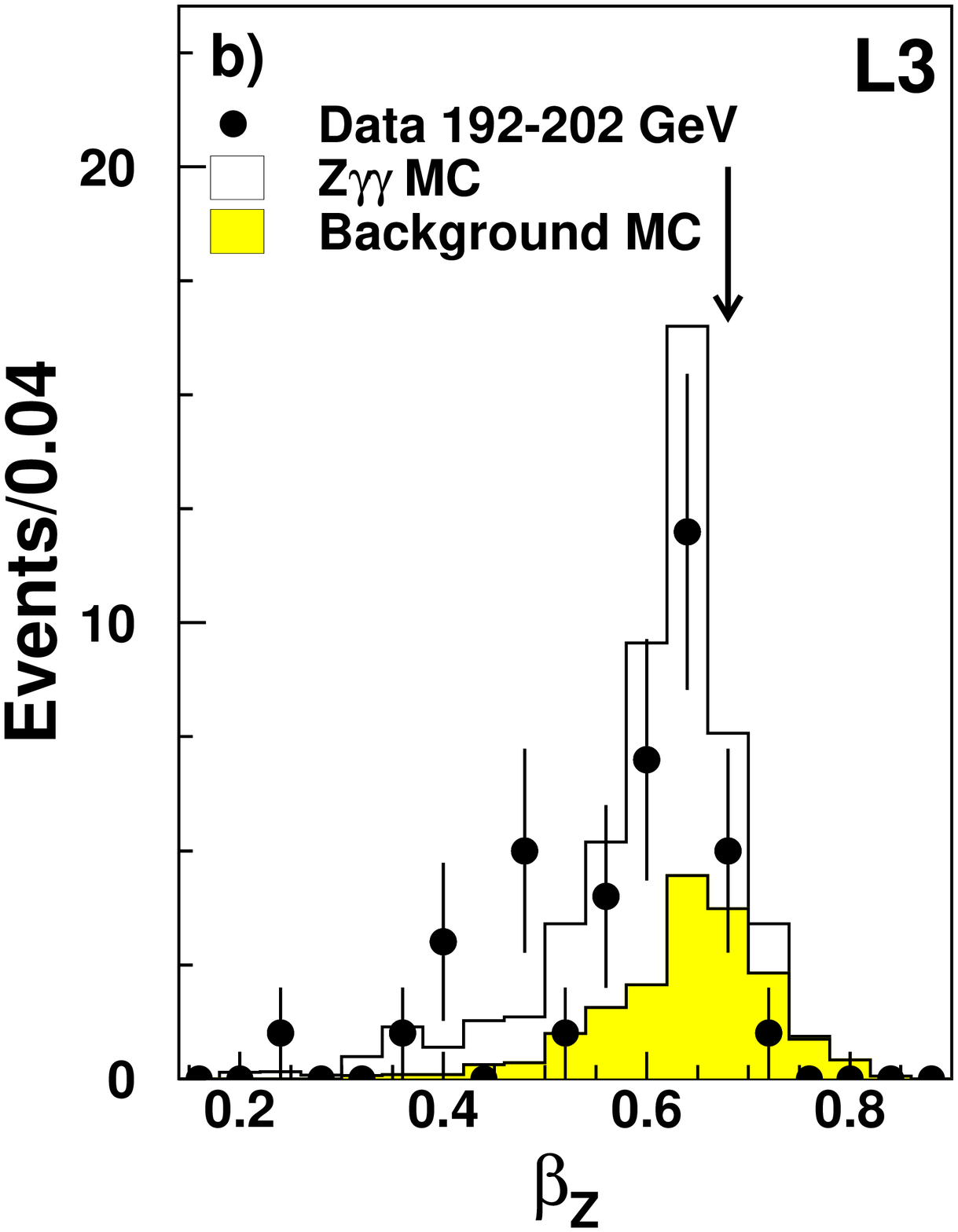}} \\
      \mbox{\includegraphics[width=.5\figwidth]{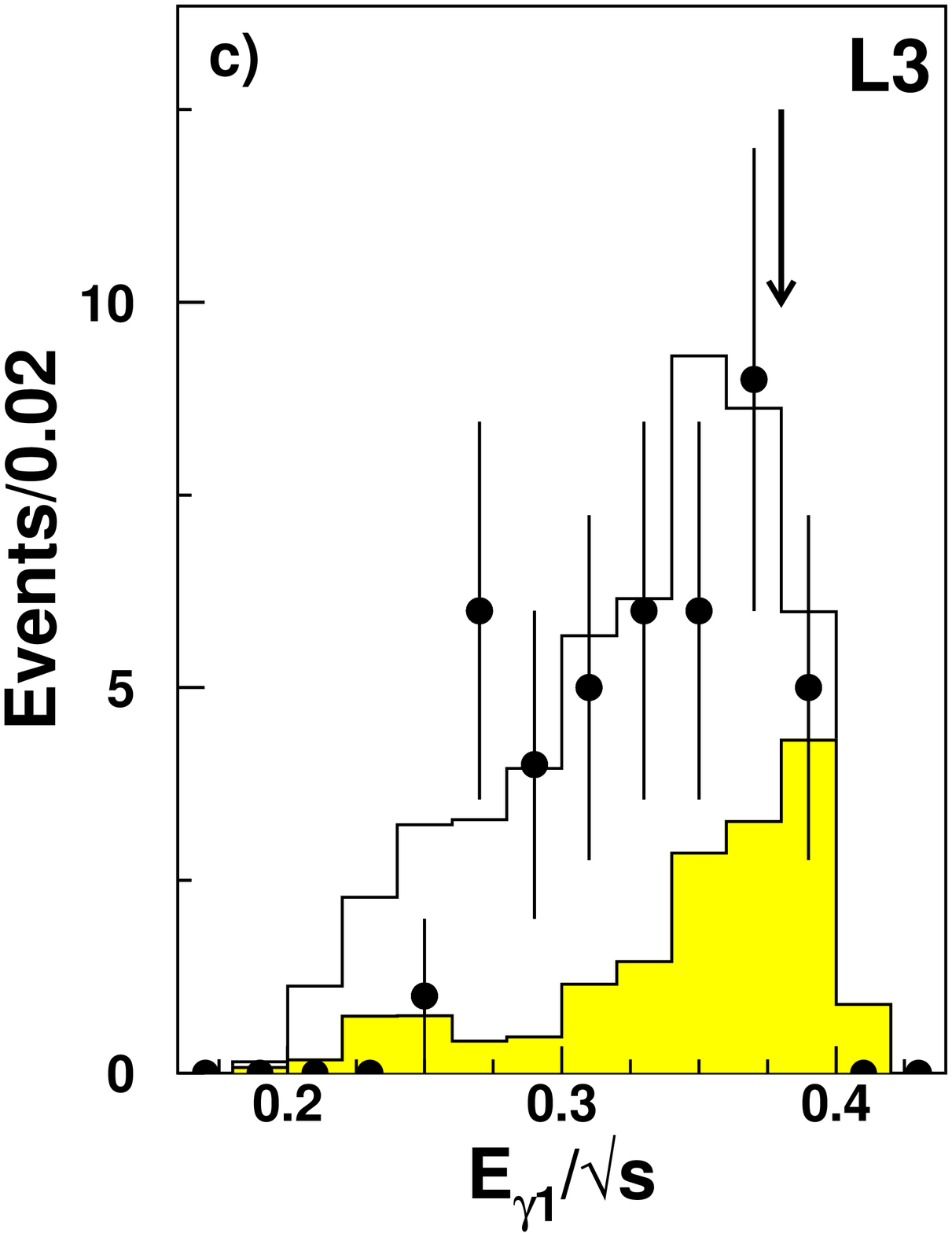}} &
      \mbox{\includegraphics[width=.5\figwidth]{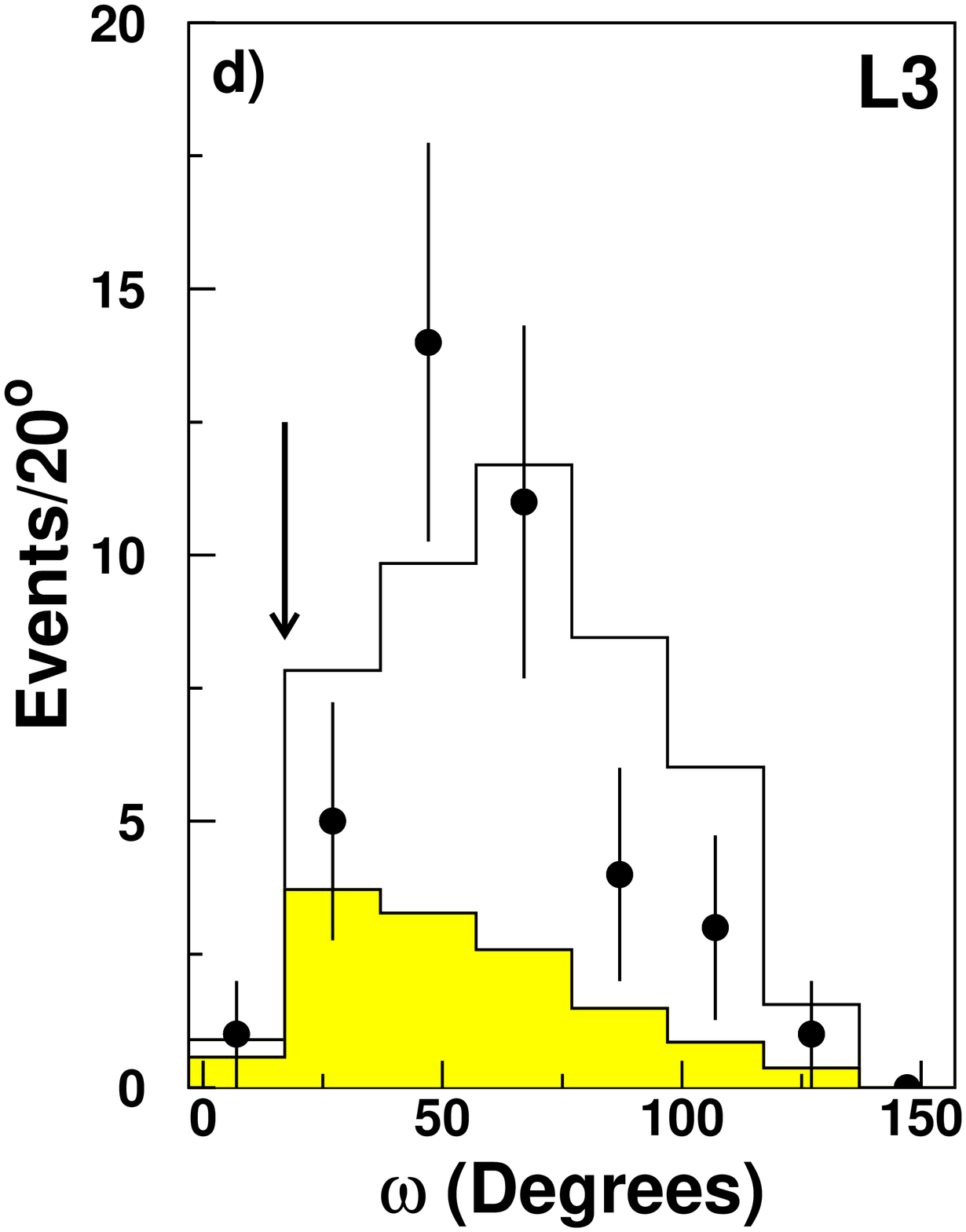}} \\
    \end{tabular}
    \icaption{Distributions of a) the  invariant mass $M_{\qqbar}$ of
    the hadronic system, b)
    the relativistic velocity 
     $\beta_{\Zo}$ of the reconstructed Z boson, c) the
    energy $E_{\gamma 1}$ of the most
    energetic photon  and d) the angle $\omega$ between the least energetic
    photon and the nearest jet. Data, signal and background Monte Carlo
    samples are shown for $\sqrt{s} = 192\GeV-202\GeV$. The
    arrows show the position of the final selection requirements. 
   In each plot, the selection criteria on the other variables are applied.
    \label{fig:1}}
  \end{center}
\end{figure}

Other classes of
background events, shown in Figure 1b
and Figure 1c,
are rejected by  an upper bound on the energy
$E_{\gamma 1}$ of the most
energetic photon. This requirement, presented in Table~\ref{sel_cuts},  
 suppresses  the resonant return to the Z,
whose photons are harder than the signal ones.
A lower bound of 17$^\circ$ on the angle $\omega$ between the
 least energetic photon and
the closest jet is also imposed.
 This requirement is more restrictive than the similar cut 
on $\cos \theta_{\rm \gamma q}$ included in the signal definition.
Data and Monte Carlo 
distributions of the selection variables are presented in
Figure~\ref{fig:1} 
for the data collected at  $\sqrt{s}$ = 192 \GeV $-$202 \GeV~when selection
criteria on all the other variables 
are applied. Good agreement between data and Monte Carlo
is observed.
\begin{table}[h]
\begin{center}
\begin{tabular}{|c|c|c|c|c|c|c|c|c|}
\hline
$\sqrt{s}$ (GeV) & 133.1  & 166.8 & 182.7 & 188.7 & 191.6 & 195.5 & 199.6 & 201.7 \\
\hline
${\rm \beta_{\rm Z}} <$  &  0.48 &  0.61 &  0.64 &   0.66 &
 0.66 &  0.67 &  0.69 &  0.70\\
${ E_{\rm \gamma_1}} (\GeV) <$ &  31.9 &  55.0 &  67.6 &  69.8  &  72.8 &  74.2 &  75.8 &  76.6 \\
\hline
\end{tabular}
\caption[3]{Energy dependent criteria for the selection of $\eeto \Zggto \qqgg$ events.}
\label{sel_cuts}
\end{center}
\end{table}
%
%
\section{Results}

The signal efficiencies and  
the  numbers of events selected in the data 
and Monte Carlo samples 
are summarised in Table~3.
The dominant  background is hadronic events with 
photons. About half of these are ``non-resonant'' events.
 In the remaining cases, 
they originate either from final state radiation or are fake photons.

\begin{figure}[p]
  \begin{center}
    \begin{tabular}{c}
      \mbox{\rotatebox{90}{
          \includegraphics[width=.73\figwidth]{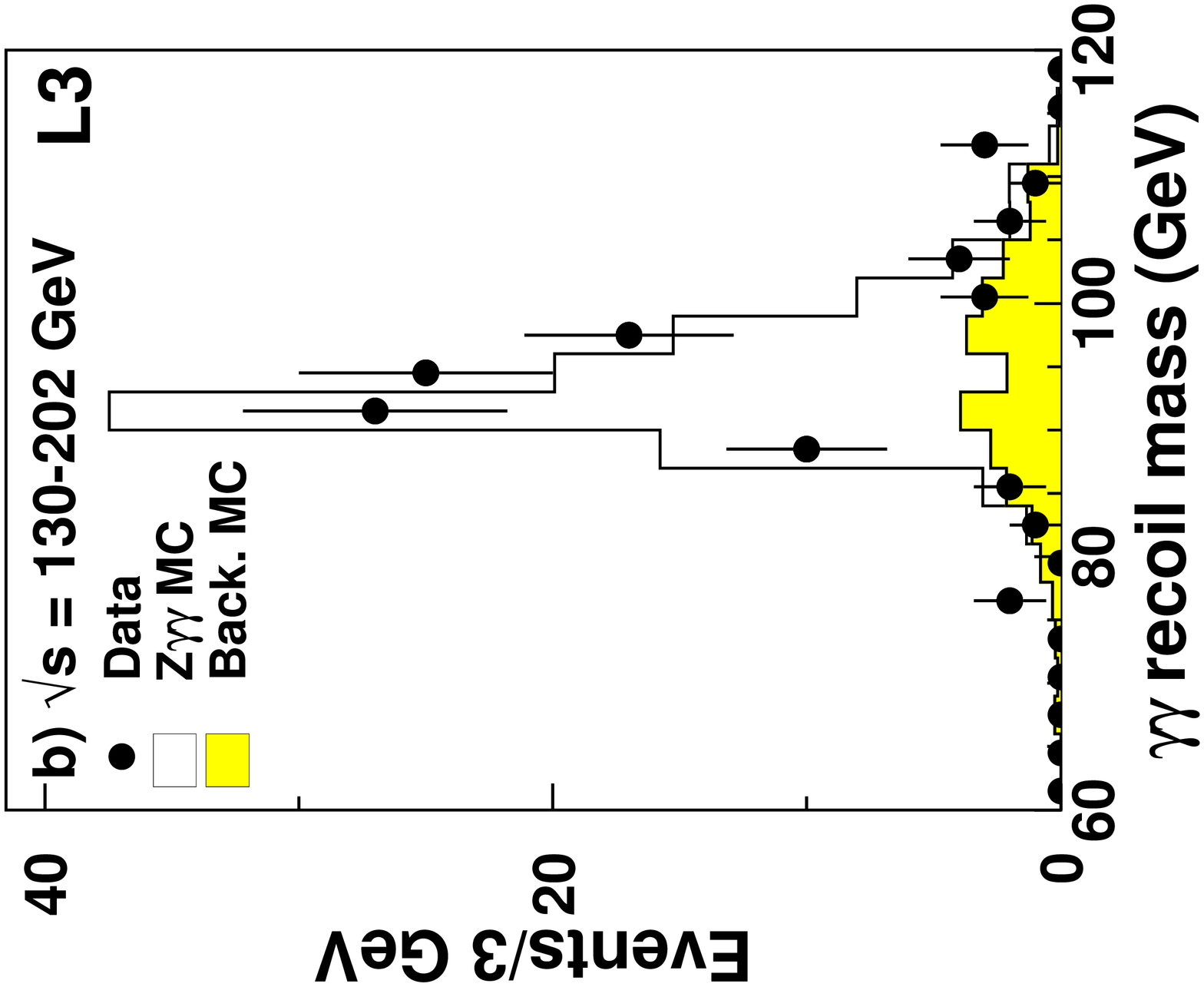}}} \\
      \mbox{\rotatebox{90}{
          \includegraphics[width=.73\figwidth]{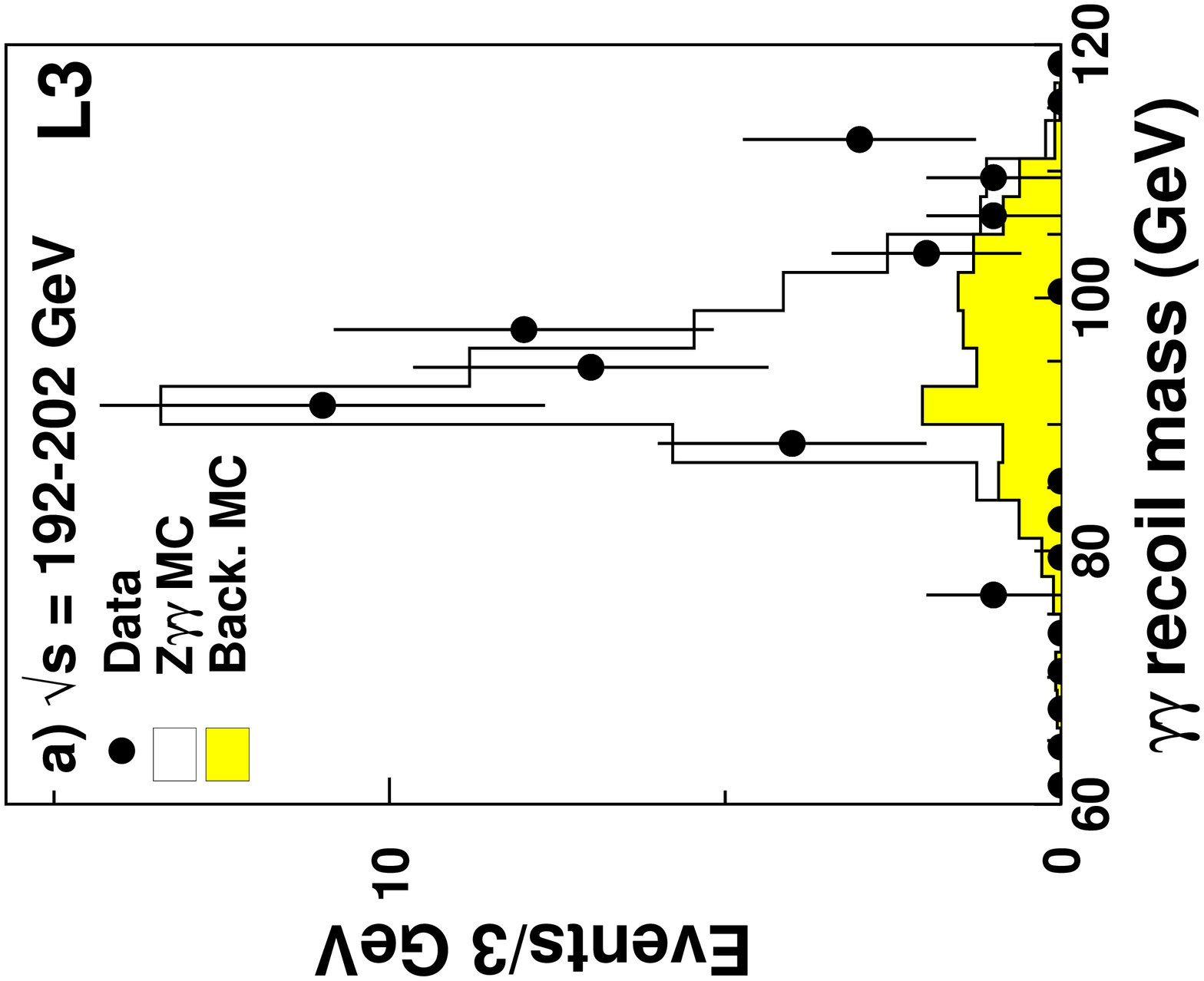}}} \\
    \end{tabular}
    \icaption{Recoil mass to the photon pairs in data, $\Zo\gamma\gamma$ and background
    Monte Carlo  for a) $\sqrt{s} = 192\GeV-202\GeV$ and
    b) the total sample.
    \label{fig:2}}
  \end{center}
\end{figure}

 A clear signal structure is observed in the spectra 
of the recoil mass to the two photons, 
as presented in Figure~\ref{fig:2} for the $\sqrt{s} = 192 \GeV - 202 \GeV$ 
data sample and for the total one.
The $\epem\ra\Zo\gamma\gamma\ra\qqbar\gamma\gamma$ cross
sections, $\sigma$,   are
determined from a fit to the corresponding spectra at each  
$\sqrt{s}$. Background predictions are fixed in the fit.
The results 
 are listed in Table \ref{tab:xs_mis}
 with  their  statistical and systematic uncertainties. 
The  systematic uncertainties  on the cross section measurement 
 are of the order of 10$\%$~\cite{l3pap}.  
The main contributions arise from the signal and background  Monte Carlo 
statistics (6$\%$) and a variation of $\pm2\%$ of the
energy scale of the hadronic 
calorimeter (6$\%$). A variation of $\pm 0.5\%$ of the energy scale of
the electromagnetic calorimeter does not yield sizable effects.
Other sources of systematic uncertainties are
the selection procedure (3$\%$) and 
the background normalisation (3$\%$). The latter is estimated by
varying by 10\% the normalisation of the ``non-resonant'' background,
as estimated from a comparison between the KK2f and PYTHIA
 Monte Carlo predictions for hadronic events with photons,
 and by 20\% that of the other backgrounds.
Uncertainties on the determination of the integrated luminosity are negligible.

\begin{figure}[p]
  \begin{center}
      \mbox{\includegraphics[width=\figwidth]{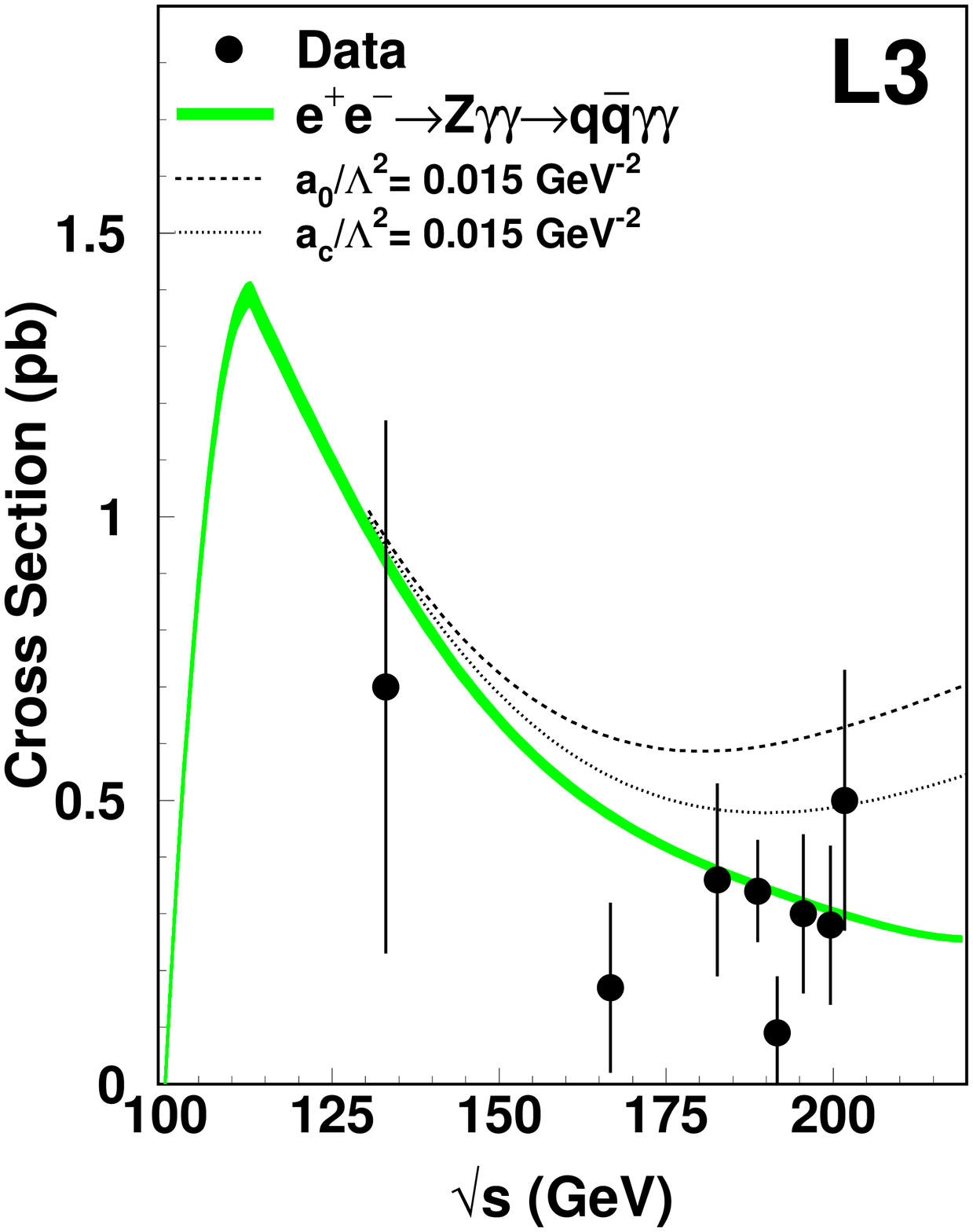}}
    \icaption{The cross section of the process 
$\epem\ra\Zo\gamma\gamma\ra\qqbar\gamma\gamma$ 
    as a function of  the centre-of-mass energy. The signal is
      defined by the phase-space cuts of Equations (1)$-$(4).
     The width of
    the band corresponds to the  Monte Carlo statistics and
    theory uncertainties. Dashed and dotted lines
    represent anomalous QGC predictions for
    $a_0/\Lambda^2$ = 0.015 ${\rm GeV^{-2}}$ and
    $a_c/\Lambda^2$ = 0.015 ${\rm GeV^{-2}}$, respectively.
    \label{fig:3}}
  \end{center}
\end{figure}

The measurements are in 
good agreement with the theoretical predictions 
$\sigma^{\rm SM}$,
as calculated with the KK2f Monte Carlo program, listed in
Table~\ref{tab:xs_mis}. 
The error on the
predictions (1.5$\%$) is the quadratic sum of
 the theory uncertainty~\cite{KK2f}
and the statistical uncertainty of the Monte Carlo sample 
generated for the calculation.
 These results are 
presented in Figure~\ref{fig:3} together with the expected evolution
with $\sqrt{s}$ of the Standard Model cross section.

 The distribution of the  recoil mass to the two photons
 for the full data sample,
 presented in Figure~\ref{fig:2}b, 
is fitted to calculate the ratio $R_{\rm \Zgg}$ 
between all the observed data and the signal expectation. 
The background predictions are fixed in the fit, which yields:
\begin{displaymath}
R_{\rm \Zgg} = \frac{\sigma}{\sigma^{\rm SM}} = 0.85 \pm 0.11 \pm 0.06
\end{displaymath}  
\noindent
in agreement with the  Standard Model.  
The first uncertainty is statistical while
 the second is systematic. The correlation 
of the energy  scale and background 
normalisation uncertainties between data samples is taken into account. \\
\begin{table}[htb]
  \begin{center}
    \begin{tabular}{|c||c|c|c||c|c|c|}
\hline
\rule{0pt}{12pt} $\sqrt{s}({\rm GeV})$  & $\varepsilon(\%)$ & $\rm Data$ & ${\rm Monte~Carlo}$ 
 & $N_s$  & $N^{\rm q\bar{q}}_b$  & $N^{Other}_b$   \\
\hline
133.1 & 45 &\phantom{0}4 &  \phantom{0}5.9 $\pm$ 0.5 & \phantom{0}5.0 $\pm$ 0.5  & 0.8 $\pm$ 0.2 & 0.08 $\pm$ 0.02\\ 
166.8 & 52 &\phantom{0}4 &  \phantom{0}6.7 $\pm$ 0.3 & \phantom{0}4.9 $\pm$ 0.3 &  1.4 $\pm$ 0.1 & 0.4\phantom{0} $\pm$ 0.1\phantom{0}\\
182.7 & 51 & 13 & 13.6 $\pm$ 0.7 & 10.8 $\pm$ 0.6 & 2.7 $\pm$ 0.2  &0.06 $\pm$ 0.02\\
188.7 & 52 & 38 & 40.3 $\pm$ 2.0 & 32.5 $\pm$ 1.7 & 7.2 $\pm$ 1.1 &0.6\phantom{0} $\pm$ 0.1\phantom{0} \\
191.6 & 42 &\phantom{0}2 & \phantom{0}5.9 $\pm$ 0.4 & \phantom{0}4.1 $\pm$ 0.3 & 1.8 $\pm$ 0.3 & 0.06 $\pm$ 0.02\\   
195.5 & 46 & 13 & 17.5  $\pm$ 0.9 & 12.4 $\pm$ 0.7 & 4.9 $\pm$ 0.5 & 0.2\phantom{0} $\pm$ 0.1\phantom{0}\\
199.6 & 46 & 14 & 15.0 $\pm$ 0.8  & 11.5 $\pm$ 0.6 &  3.4 $\pm$ 0.5 & 0.13 $\pm$ 0.05 \\
201.7 & 48 & \phantom{0}9 & \phantom{0}6.9 $\pm$ 0.5 & \phantom{0}5.2  $\pm$ 0.4 & 1.7 $\pm$ 0.3 & 0.06 $\pm$ 0.02\\
\hline
    \end{tabular}
    \icaption[tab:1]{Yields of the
      $\epem\ra\Zo\gamma\gamma\ra\qqbar\gamma\gamma$ selection. The signal
      efficiencies $\varepsilon$ are given,
 together with the observed and expected 
      numbers of events. The right half of the table 
details the composition of the
      Monte Carlo samples with  $N_s$ denoting the signal,
      $N^{\rm q\bar{q}}_b$  the $\rm{q\bar{q}}$ and  $N^{Other}_b$ the other
      backgrounds. The uncertainties are statistical only.}   
  \end{center}
\end{table}
\begin{table}[htb] 
\begin{center}
\begin{tabular}{|c|c|c|}
\hline
\rule{0pt}{12pt} $~~~\sqrt{s}~{\rm (GeV)}~~~$ & $~~~\sigma
$~({\rm pb})~~~ & $~~~\sigma^{\rm SM}
$~({\rm pb})~~~\\
\hline
133.1    & $0.70 \pm 0.40 \pm 0.07$  & $0.923 \pm 0.012$\\
166.8    & $0.17 \pm 0.13 \pm 0.02$ &$ 0.475 \pm  0.006$\\   
182.7    & $0.36 \pm 0.13 \pm 0.04$ & $  0.379 \pm  0.004$\\
188.7    & $0.34 \pm 0.06  \pm 0.03$ & $  0.350 \pm 0.004$ \\
191.6    & $0.09 \pm 0.09  \pm 0.01 $ & $  0.326 \pm  0.004$ \\
195.5    & $0.30 \pm 0.11  \pm 0.03$  & $ 0.321 \pm  0.004$\\
199.6    & $0.28 \pm 0.11  \pm 0.03$ & $ 0.304 \pm 0.004 $\\
201.7    & $0.50 \pm 0.18  \pm 0.05$  & $    0.296\pm  0.003$\\
\hline
\end{tabular}
\icaption{Results of the measurements 
of the ${\rm e^+e^- \rightarrow Z \gamma \gamma \rightarrow \qqgg}$  cross section, 
$\sigma$, with
  statistical and systematic uncertainties.
 The predicted values of cross sections,
  $\sigma^{\rm SM}$,
    are also listed.}
\label{tab:xs_mis}
\end{center}
\end{table} 

%
%
\section{Study of Quartic Gauge Boson Couplings}

The contribution of anomalous QGCs to  $\Zo\gamma\gamma$ production
 is described by two additional dimension-six  terms in the electroweak Lagrangian~\cite{bb1,sw}:
\begin{eqnarray*}
{\cal L}^0_6 & = &  -{\pi\alpha \over 4\Lambda^2} a_0 F_{\mu\nu}F^{\mu\nu}
\vec{W}_\rho\cdot\vec{W}^\rho\\
{\cal L}^c_6 & = &  -{\pi\alpha \over 4\Lambda^2} a_c F_{\mu\rho}F^{\mu\sigma}
\vec{W}^\rho\cdot\vec{W}_\sigma,
\end{eqnarray*}
where $\alpha$ is the fine structure constant, $F_{\mu\nu}$ is the
field strength tensor of the photon and $\vec{W}_\sigma$ is the 
 weak boson
field. 
The parameters $a_0$ and $a_c$ describe the  strength of the QGCs and
$\Lambda$ represents the unknown scale of the New Physics responsible for the
anomalous contributions. In the Standard Model, $a_0 = a_c = 0$. 
A more detailed description of  QGCs has recently appeared~\cite{bb2}.
Indirect limits on  QGCs were derived from precision
measurements at the Z pole~\cite{eboli}.

\begin{figure}[p]
  \begin{center}
          \mbox{\includegraphics[width=1.\figwidth]{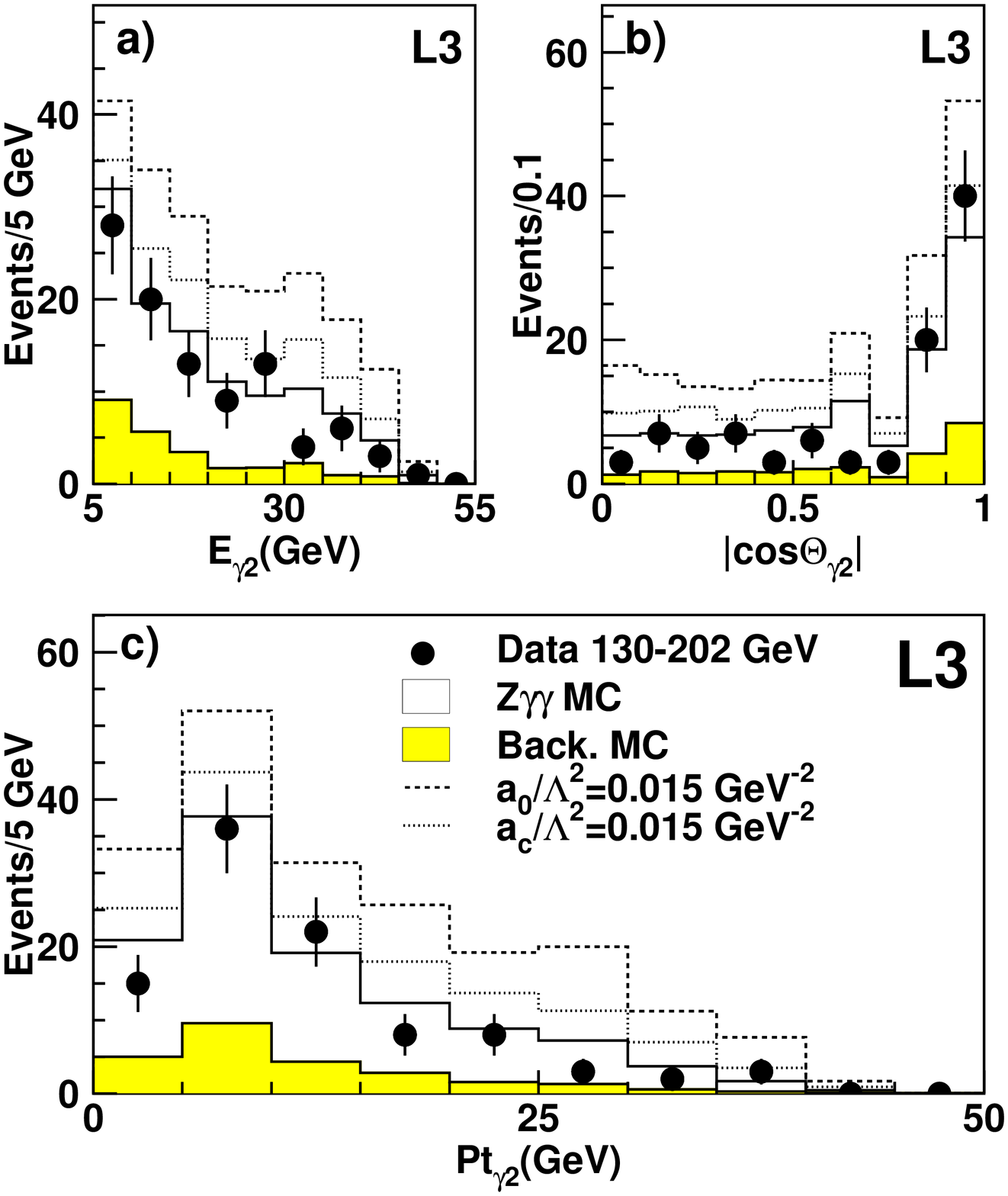}} \\
          \icaption{Distributions for the least energetic photon.
  a) the energy $E_{\gamma 2}$, 
  b) the cosine of its polar angle $|\cos{\theta_{\gamma 2}}|$,
c) its transverse momentum $Pt_{\gamma 2}$ 
with respect to the beam axis. 
 Data,  signal and background Monte Carlo 
    are displayed for the full data sample together with QGC predictions.
      \label{fig:dev2}}
  \end{center}
\end{figure}

\begin{figure}[hbt]
  \begin{center}
      \mbox{\includegraphics[width=1.\figwidth]{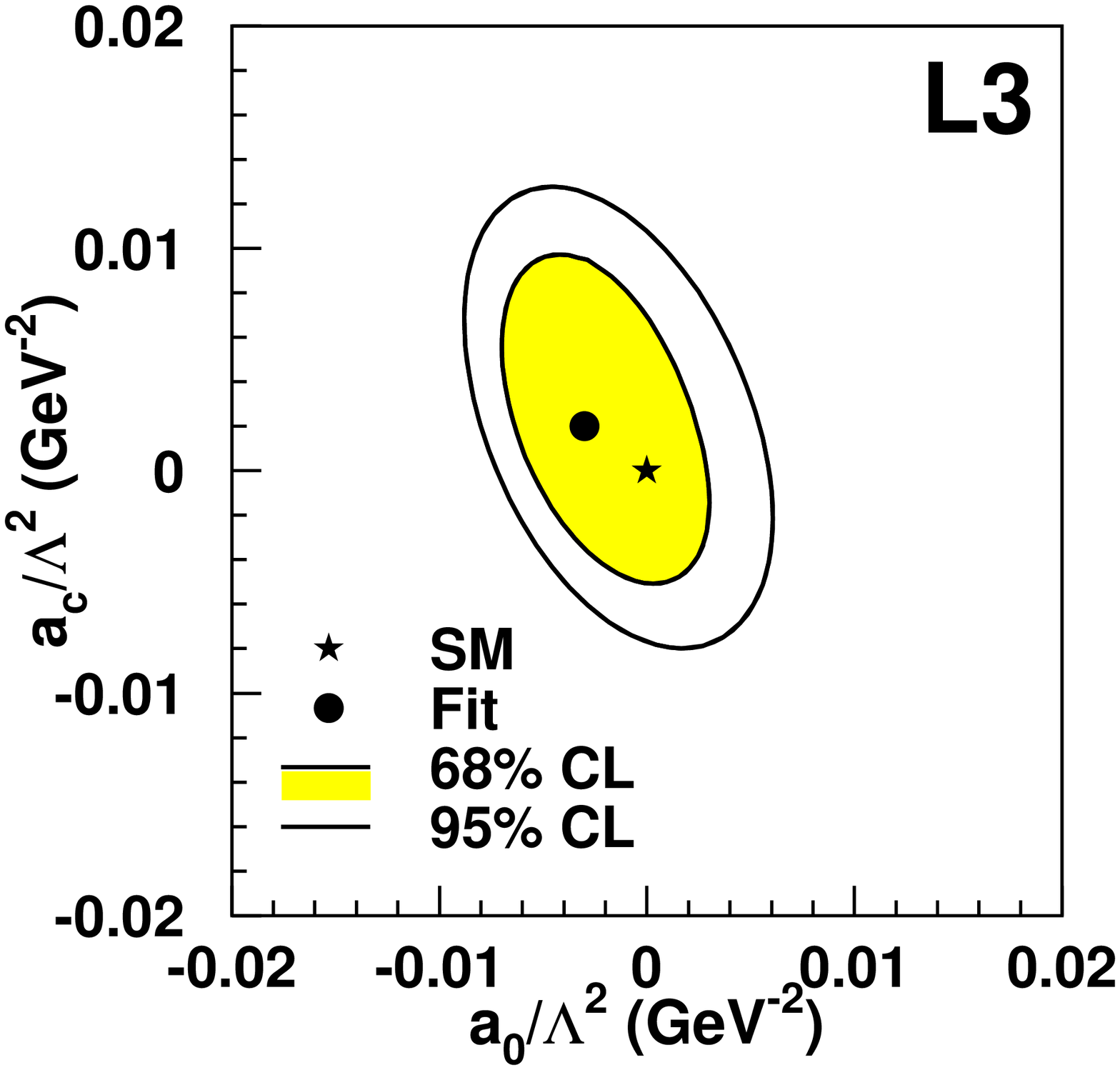}}
    \icaption{ Two dimensional
    contours for the QGC parameters $a_0/\Lambda^2$ and 
    $a_c/\Lambda^2$. 
The fit result is shown together with the Standard Model (SM) predictions.
      \label{fig:5}}
  \end{center}
\end{figure}

Anomalous  values of QGCs are expected to  manifest 
themselves via deviations in the total $\epem\ra\Zo\gamma\gamma$ 
cross section, as  presented in Figure~\ref{fig:3}.
In the Standard Model, Z$\gamma\gamma$ production occurs via bremsstrahlung
with the low energy photon preferentially produced close to the beam
direction.
The QGC $s$-channel
production  results instead in a  harder energy spectrum 
and  a more central angular distribution of the
least energetic photon~\cite{sw}. 
Distributions for this photon
of the reconstructed 
energy, the  cosine of the polar angle and the transverse momentum for the full data sample
are compared in
Figure \ref{fig:dev2}
with the predictions from signal and background  Monte Carlo.
Predictions in the case of a non zero 
value of $a_0/\Lambda^2$ or $a_c/\Lambda^2$ are also shown. 
They are obtained by reweighting~\cite{l3pap} 
the Standard Model signal 
Monte Carlo events with an analytical 
calculation of the QGC matrix element~\cite{sw}.
Monte Carlo studies indicate the transverse momentum as the most sensitive distribution 
to possible anomalous QGC contributions.
A fit to this distribution is performed for each data sample, leaving one of
the two QGCs free at a time and fixing the other to zero.  
It yields the 68\% confidence level results:
\begin{displaymath}
a_0/\Lambda^2  =  -0.002^{+ 0.003}_{-0.002} \GeV^{-2}\,\,\,{\rm and}\,\,\,
a_c/\Lambda^2  =  -0.001^{+ 0.006}_{-0.004} \GeV^{-2}\,,
\end{displaymath}
in agreement with the expected Standard Model values of zero. 
A   simultaneous fit to both the parameters  gives
the 95\% confidence level limits:
\begin{displaymath}
-0.008\GeV^{-2} < a_0/\Lambda^2  < 0.005\GeV^{-2}\,\,\,{\rm and}\,\,\,
-0.007 \GeV^{-2}< a_c/\Lambda^2  < 0.011\GeV^{-2}\,,
\end{displaymath}
as shown in Figure~\ref{fig:5}.
A correlation coefficient of $-57\%$ is observed. The experimental systematic uncertainties and those
 on the Standard  Model
$\epem\ra\Zo\gamma\gamma\ra\qqbar\gamma\gamma$ 
cross section predictions are taken into account in the fit.

%
%
\section*{Acknowledgements}

We wish to express our gratitude to the CERN accelerator divisions for the
superb performance and the continuous and successful upgrade of the
LEP machine.  
We acknowledge the contributions of the engineers  and technicians who
have participated in the construction and maintenance of this experiment.

\section*{Appendix}

To allow the combination of our results 
with those of the other LEP experiments,
the cross sections $\sigma$ are also
measured in the more restrictive phase space obtained by modifying the conditions~(2) and (4) into 
$|\cos{\theta_{\gamma}}|< 0.95$ and $\cos{\theta_{\rm \gamma q}}< 0.9$, respectively.
The results are:
\begin{center}\begin{tabular}{rcl}
$\rm \sigma(182.7 { \GeV}) $&=&$ \rm0.11 \pm 0.11 \pm 0.01~pb~~~(SM:0.233 \pm  0.003~pb)$ \\
$\rm \sigma(188.7 { \GeV}) $&=&$ \rm0.28 \pm 0.07 \pm 0.03~pb~~~(SM:0.214 \pm 0.003~pb)$\\
$\rm \sigma(194.5 { \GeV}) $&=&$ \rm0.15 \pm 0.07 \pm 0.02~pb~~~(SM:0.197 \pm  0.003~pb)$\\
$\rm \sigma(200.2 { \GeV}) $&=&$ \rm0.15 \pm 0.07 \pm 0.01~pb~~~(SM:0.185 \pm 0.003~pb)$.\\
\end{tabular}\end{center}
The first uncertainty is statistical, the second systematic and the values
 in parentheses indicate the Standard Model predictions. 
The samples at $\sqrt{s} = 192\GeV - 196\GeV$
 and $\sqrt{s} = 200\GeV - 202\GeV$ 
are respectively merged into the
 $\sqrt{s} = 194.5\GeV$ and $\sqrt{s} = 200.2\GeV$ ones.

\newpage
%
%

\bibliographystyle{l3stylem}

%
%

\newpage
\typeout{   }     
\typeout{Using author list for paper 234 only }
\typeout{$Modified: Fri Jan 26 2001 by smele $}
\typeout{!!!!  This should only be used with document option a4p!!!!}
\typeout{   }
%
%
%
%
%
%

\newcount\tutecount  \tutecount=0
\def\tutenum#1{\global\advance\tutecount by 1 \xdef#1{\the\tutecount}}
\def\tute#1{$^{#1}$}
\tutenum\aachen            
\tutenum\nikhef            
\tutenum\mich              
\tutenum\lapp              
\tutenum\basel             
\tutenum\lsu               
\tutenum\beijing           
\tutenum\berlin            
\tutenum\bologna           
\tutenum\tata              
\tutenum\ne                
\tutenum\bucharest         
\tutenum\budapest          
\tutenum\mit               
\tutenum\debrecen          
\tutenum\florence          
\tutenum\cern              
\tutenum\wl                
\tutenum\geneva            
\tutenum\hefei             
\tutenum\lausanne          
\tutenum\lecce             
\tutenum\lyon              
\tutenum\madrid            
\tutenum\milan             
\tutenum\moscow            
\tutenum\naples            
\tutenum\cyprus            
\tutenum\nymegen           
\tutenum\caltech           
\tutenum\perugia           
\tutenum\peters            
\tutenum\cmu               
\tutenum\potenza           
\tutenum\prince            
\tutenum\riverside         
\tutenum\rome              
\tutenum\salerno           
\tutenum\ucsd              
\tutenum\sofia             
\tutenum\korea             
\tutenum\alabama           
\tutenum\utrecht           
\tutenum\purdue            
\tutenum\psinst            
\tutenum\zeuthen           
\tutenum\eth               
\tutenum\hamburg           
\tutenum\taiwan            
\tutenum\tsinghua          

{
\parskip=0pt
\noindent
{\bf The L3 Collaboration:}
\ifx\selectfont\undefined
 \baselineskip=10.8pt
 \baselineskip\baselinestretch\baselineskip
 \normalbaselineskip\baselineskip
 \ixpt
\else
 \fontsize{9}{10.8pt}\selectfont
\fi
\medskip
\tolerance=10000
\hbadness=5000
\raggedright
\hsize=162truemm\hoffset=0mm
\def\r{\rlap,}
\noindent

M.Acciarri\r\tute\milan\
P.Achard\r\tute\geneva\ 
O.Adriani\r\tute{\florence}\ 
M.Aguilar-Benitez\r\tute\madrid\ 
J.Alcaraz\r\tute\madrid\ 
G.Alemanni\r\tute\lausanne\
J.Allaby\r\tute\cern\
A.Aloisio\r\tute\naples\ 
M.G.Alviggi\r\tute\naples\
G.Ambrosi\r\tute\geneva\
% ...
\newpage
\begin{list}{A}{\itemsep=0pt plus 0pt minus 0pt\parsep=0pt plus 0pt minus 0pt
                \topsep=0pt plus 0pt minus 0pt}
\item[\aachen]
 I. Physikalisches Institut, RWTH, D-52056 Aachen, FRG$^{\S}$\\
 III. Physikalisches Institut, RWTH, D-52056 Aachen, FRG$^{\S}$
\item[\nikhef] National Institute for High Energy Physics, NIKHEF, 
     and University of Amsterdam, NL-1009 DB Amsterdam, The Netherlands
\item[\mich] University of Michigan, Ann Arbor, MI 48109, USA
\item[\lapp] Laboratoire d'Annecy-le-Vieux de Physique des Particules, 
     LAPP,IN2P3-CNRS, BP 110, F-74941 Annecy-le-Vieux CEDEX, France
\item[\basel] Institute of Physics, University of Basel, CH-4056 Basel,
     Switzerland
\item[\lsu] Louisiana State University, Baton Rouge, LA 70803, USA
\item[\beijing] Institute of High Energy Physics, IHEP, 
  100039 Beijing, China$^{\triangle}$ 
\item[\berlin] Humboldt University, D-10099 Berlin, FRG$^{\S}$
\item[\bologna] University of Bologna and INFN-Sezione di Bologna, 
     I-40126 Bologna, Italy
\item[\tata] Tata Institute of Fundamental Research, Bombay 400 005, India
\item[\ne] Northeastern University, Boston, MA 02115, USA
\item[\bucharest] Institute of Atomic Physics and University of Bucharest,
     R-76900 Bucharest, Romania
\item[\budapest] Central Research Institute for Physics of the 
     Hungarian Academy of Sciences, H-1525 Budapest 114, Hungary$^{\ddag}$
\item[\mit] Massachusetts Institute of Technology, Cambridge, MA 02139, USA
\item[\debrecen] KLTE-ATOMKI, H-4010 Debrecen, Hungary$^\P$
\item[\florence] INFN Sezione di Firenze and University of Florence, 
     I-50125 Florence, Italy
\item[\cern] European Laboratory for Particle Physics, CERN, 
     CH-1211 Geneva 23, Switzerland
\item[\wl] World Laboratory, FBLJA  Project, CH-1211 Geneva 23, Switzerland
\item[\geneva] University of Geneva, CH-1211 Geneva 4, Switzerland
\item[\hefei] Chinese University of Science and Technology, USTC,
      Hefei, Anhui 230 029, China$^{\triangle}$
\item[\lausanne] University of Lausanne, CH-1015 Lausanne, Switzerland
\item[\lecce] INFN-Sezione di Lecce and Universit\`a Degli Studi di Lecce,
     I-73100 Lecce, Italy
\item[\lyon] Institut de Physique Nucl\'eaire de Lyon, 
     IN2P3-CNRS,Universit\'e Claude Bernard, 
     F-69622 Villeurbanne, France
\item[\madrid] Centro de Investigaciones Energ{\'e}ticas, 
     Medioambientales y Tecnolog{\'\i}cas, CIEMAT, E-28040 Madrid,
     Spain${\flat}$ 
\item[\milan] INFN-Sezione di Milano, I-20133 Milan, Italy
\item[\moscow] Institute of Theoretical and Experimental Physics, ITEP, 
     Moscow, Russia
\item[\naples] INFN-Sezione di Napoli and University of Naples, 
     I-80125 Naples, Italy
\item[\cyprus] Department of Natural Sciences, University of Cyprus,
     Nicosia, Cyprus
\item[\nymegen] University of Nijmegen and NIKHEF, 
     NL-6525 ED Nijmegen, The Netherlands
\item[\caltech] California Institute of Technology, Pasadena, CA 91125, USA
\item[\perugia] INFN-Sezione di Perugia and Universit\`a Degli 
     Studi di Perugia, I-06100 Perugia, Italy   
\item[\peters] Nuclear Physics Institute, St. Petersburg, Russia
\item[\cmu] Carnegie Mellon University, Pittsburgh, PA 15213, USA
\item[\potenza] INFN-Sezione di Napoli and University of Potenza, 
     I-85100 Potenza, Italy
\item[\prince] Princeton University, Princeton, NJ 08544, USA
\item[\riverside] University of Californa, Riverside, CA 92521, USA
\item[\rome] INFN-Sezione di Roma and University of Rome, ``La Sapienza",
     I-00185 Rome, Italy
\item[\salerno] University and INFN, Salerno, I-84100 Salerno, Italy
\item[\ucsd] University of California, San Diego, CA 92093, USA
\item[\sofia] Bulgarian Academy of Sciences, Central Lab.~of 
     Mechatronics and Instrumentation, BU-1113 Sofia, Bulgaria
\item[\korea]  Laboratory of High Energy Physics, 
     Kyungpook National University, 702-701 Taegu, Republic of Korea
\item[\alabama] University of Alabama, Tuscaloosa, AL 35486, USA
\item[\utrecht] Utrecht University and NIKHEF, NL-3584 CB Utrecht, 
     The Netherlands
\item[\purdue] Purdue University, West Lafayette, IN 47907, USA
\item[\psinst] Paul Scherrer Institut, PSI, CH-5232 Villigen, Switzerland
\item[\zeuthen] DESY, D-15738 Zeuthen, 
     FRG
\item[\eth] Eidgen\"ossische Technische Hochschule, ETH Z\"urich,
     CH-8093 Z\"urich, Switzerland
\item[\hamburg] University of Hamburg, D-22761 Hamburg, FRG
\item[\taiwan] National Central University, Chung-Li, Taiwan, China
\item[\tsinghua] Department of Physics, National Tsing Hua University,
      Taiwan, China
\item[\S]  Supported by the German Bundesministerium 
        f\"ur Bildung, Wissenschaft, Forschung und Technologie
\item[\ddag] Supported by the Hungarian OTKA fund under contract
numbers T019181, F023259 and T024011.
\item[\P] Also supported by the Hungarian OTKA fund under contract
  numbers T22238 and T026178.
\item[$\flat$] Supported also by the Comisi\'on Interministerial de Ciencia y 
        Tecnolog{\'\i}a.
\item[$\sharp$] Also supported by CONICET and Universidad Nacional de La Plata,
        CC 67, 1900 La Plata, Argentina.
\item[$\diamondsuit$] Also supported by Panjab University, Chandigarh-160014, 
        India.
\item[$\triangle$] Supported by the National Natural Science
  Foundation of China.
\end{list}
}
\vfill


\newpage

%
%

\end{document}